\documentclass[twocolumn,secnumarabic,amssymb, amsmath, nofootinbib,tightenlines,
nobibnotes, aps, prl,epsfig]{revtex4}
\usepackage{graphicx}
\usepackage{dcolumn}
\usepackage{bm}
\begin{document}
\preprint{APS/123-QED}
\title{Evolution of the color dipole cross section}

\author{G.R.Boroun}%
 \email{ boroun@razi.ac.ir }
\affiliation{ Physics Department, Razi University, Kermanshah
67149, Iran}

\date{\today}
\begin{abstract}
Using Laplace transform techniques, we describe the evolution of
the color dipole cross section, at the leading-order and
next-to-leading order approximations, from the
Bartels-Golec-Biernat-Kowalski model in a kinematical region of
low values of the Bjorken variable $x$ and a wide range of
transverse dipole size $r$. This evolution method  shows that the
saturation scale and geometric scaling are retained. We derived
analytical results for the integrated and unintegrated color
dipole gluon distribution functions and compared them with the
CJ15 parametrization group and the unintegrated color dipole gluon
distribution models respectively. The Sudakov form factor into the
evolution of the unintegrated color dipole gluon distribution is
incorporated and the results are considered at small and large
values of
$k_{t}^{2}$.\\

\end{abstract}
 \pacs{***}
\keywords{****} 
\maketitle
\subsection{1. Introduction}

In the color dipole model (CDM), the virtual photon $\gamma^{*}$
exchanged between the electron and proton currents with virtuality
$Q^2$, split into a quark-antiquark pair (a dipole) which then
interacts with the target proton via gluon exchanges \cite{Ref1}.
The quark and antiquark in this dipole carry a fraction $z$ and
$1-z$ of the photon longitudinal momentum respectively. The
transverse size between the quark and antiquark is given by the
vector $\mathbf{r}$. The interaction of small- sized color dipoles
with nucleons is governed by perturbative QCD (pQCD) and also the
basic dipole-factorization holds at large dipole sizes
\cite{Ref2}. The total $\gamma^{*}p$ cross section is given by
\begin{eqnarray}
\sigma_{L,T}^{\gamma^{*}p}(x,Q^{2})=\sum_{f}\int d^{2}\mathbf{r}
\int_{0}^{1} dz
|\Psi_{L,T}(\mathbf{r},z;Q^{2})|^{2}\sigma_{\mathrm{dip}}({x},\mathbf{r}),
\end{eqnarray}
where the sum over quark flavours f is performed. Here $x$ is the
Bjorken scaling and $\Psi_{L,T}(\mathbf{r},z;Q^2)$ are the
appropriate spin averaged light-cone wave functions of the photon,
which give the probability for the occurrence of a
$(q\overline{q})$ fluctuation of transverse size with respect to
the photon polarization \cite{Ref3}. The photon wave function
depends on the mass of the quarks in the $q\overline{q}$ dipole,
therefore contributions  depend on the mass of the quarks by
modifying the Bjorken variable $x$ in the dipole cross section as
 $x{\rightarrow}\widetilde{x}_{f}{\equiv}x(1+4m_{c}^{2}/Q^{2})$ with the
number of active flavor $n_{f}=4$ where $m_{c}$ is the mass of the
quark of charm (with $m_{c}=1.4~\mathrm{GeV}$).\\
The dipole cross-section, $\sigma_{\mathrm{dip}}({x},r)$, which
related to the imaginary part of the $(q\overline{q})p$ forward
scattering amplitude, in the Golec-Biernat-Wusthof (GBW) model
\cite{Ref4} is defined by the following form
\begin{eqnarray}
\sigma_{\mathrm{dip}}(x,\mathbf{r})=\sigma_{0}\bigg{\{}1-
\exp\bigg{(}-r^2Q_{\mathrm{sat}}^2(x)/4\bigg{)} \bigg{\}},
\end{eqnarray}
where $Q_{\mathrm{sat}}(x)$ plays the role of the saturation by
the form $Q_{\mathrm{sat}}^2(x)=Q_{0}^{2}(x_{0}/x)^\lambda$
($Q_{0}^{2}=1~\mathrm{GeV}^2$) and it becomes a function of a
single variable $rQ_{s}$ (for all values of $r$ and $x$) as has a
property of geometric scaling \cite{Ref5} by the following form
\begin{eqnarray}
\sigma_{\mathrm{dip}}(x,{r})=\sigma_{\mathrm{dip}}({r}Q_{s}(x)).
\end{eqnarray}
Indeed, the GBW model is used to describe the inclusive DIS data
within a certain approximation. As is well known, this model is
reasonable only at small transverse momenta $k_{t}$ (after the
Fourier transform), as its exponential decay contradicts the
expected perturbative behavior at large $k_{t}$. The
Bartels-Golec-Biernat-Kowalski (BGK) model \cite{Ref6}, is another
phenomenological approach to dipole cross section and reads
\begin{eqnarray}
\sigma_{\mathrm{dip}}(x,\mathbf{r})=\sigma_{0}\bigg{\{}1-
\exp\bigg{(}
\frac{-\pi^2r^2\alpha_{s}(\mu^2)xg(x,\mu^2)}{N_{c}~\sigma_{0}}
\bigg{)} \bigg{\}},
\end{eqnarray}
where $g(x,\mu^2)$ is the gluon collinear PDF.  The dipole cross
section in the BGK model is expanded into the DGLAP improved
saturation model. In this paper we wish to evolve the dipole cross
section directly into $x$ and $r$ using a Laplace transform
technique and obtain an analytical method for the solution of the
dipole cross section in terms of  known initial gluon
distribution. Then, we apply this analytical function to test the
consistency of the saturation scale. We present unintegrated gluon
distribution with and without Sudakov form factors from the
evolution of the gluon density, which gives a good description of
the gluon distribution owing to the parametrization methods.\\

\subsection{2. Method}

The color dipole cross section in the BGK model \cite{Ref6} reads
\begin{eqnarray}
\sigma_{\mathrm{dip}}(x,r)=\sigma_{0}\{1-\exp(-\frac{\sigma_{\mathrm{DGLAP}}({x},r)}{\sigma_{0}})\},
\end{eqnarray}
where
\begin{eqnarray}
\sigma_{\mathrm{DGLAP}}({x},r)=\frac{\pi^{2}}{N_{c}}r^{2}\alpha_{s}(\mu^{2})xg({x},\mu^{2}),
\end{eqnarray}
with $N_{c}=3$ \cite{Ref7, Ref2}. The hard scale, $\mu^{2}$, is
assumed to be
\begin{eqnarray}
\mu^{2}=C/r^{2}+\mu^{2}_{0},
\end{eqnarray}
where the parameters $C$ and $\mu^{2}_{0}$ are obtained from the
fit to the DIS data as summarized in \cite{Ref7}. The gluon
distribution $xg(x,\mu^{2})$ is dominant at low $x$ in  the DGLAP
evolution equations \cite{Ref8, Ref9, Ref10} as the gluonic sector
is
\begin{eqnarray}
\frac{{\partial}G(x,\mu^{2})}{{\partial}{\ln}\mu^{2}}{\simeq}{\int_{x}^{1}}\frac{x}{y^2}dyP_{gg}(\frac{x}{y},\alpha_{s}(\mu^{2}))G(y,\mu^{2}),
\end{eqnarray}
where $G(x,\mu^{2})=xg(x,\mu^{2})$. The splitting function
$P_{gg}(x,\alpha_{s})$ at the higher-order corrections reads
\begin{eqnarray}
P_{gg}(x,
\alpha_{s})&=&\frac{\alpha_{s}}{2\pi}P^{\mathrm{LO}}_{gg}(x,
\alpha_{s})+(\frac{\alpha_{s}}{2\pi})^{2}P^{\mathrm{NLO}}_{gg}(x,
\alpha_{s}),
\end{eqnarray}
where the running coupling is defined by the following form in the
leading-order (LO) renormalization group equation,
\begin{eqnarray}
\mu^{2}\frac{d\alpha_{s}(\mu^2)}{d\mu^2}=-\bigg{(}\frac{11C_{A}-4T_{R}n_{f}}{12\pi}\bigg{)}\alpha^{2}_{s}(\mu^2)
\end{eqnarray}
with $C_{A}=N_{c}$,
 $T_{R}=\frac{1}{2}$ and $n_{f}=4$ (the active quark flavor
 number).\\
 Substituting Eq.(6) into (8), we can rewrite the evolution of $\sigma_{\mathrm{DGLAP}}({x},r)$
 as
\begin{eqnarray}
\frac{{\partial}\sigma_{\mathrm{DGLAP}}(x,\mu^{2})}{{\partial}{\ln}\mu^{2}}=-\alpha_{s}(\mu^{2})r^2
\frac{{\partial}}{{\partial}{\ln}\mu^{2}}\bigg{(}\frac{1}{\alpha_{s}(\mu^{2})r^2}
\bigg{)}{\times}~~~~~~~~~~~\nonumber\\
\sigma_{\mathrm{DGLAP}}(x,\mu^{2})+{\int_{x}^{1}}\frac{x}{y^2}dyP_{gg}(\frac{x}{y},\alpha_{s}(\mu^{2}))\sigma_{\mathrm{DGLAP}}(x,\mu^{2}).~~
\end{eqnarray}
By writing
\begin{eqnarray}
d\ln(\mu^2)=-\frac{2C}{r^3\mu^2}dr,
\end{eqnarray}
we can rewrite Eq.(11) in the following form
\begin{eqnarray}
d\sigma_{\mathrm{DGLAP}}(x,r)=\bigg{[}\frac{2}{r}+\frac{d{\ln}\alpha_{s}(r)}{dr}\bigg{]}{dr}\sigma_{\mathrm{DGLAP}}(x,r)~~~~~~~~~~\nonumber\\
-\frac{2C}{r^3\bigg{(}\frac{C}{r^2}+\mu_{0}^{2}\bigg{)}}dr{\int_{x}^{1}}\frac{x}{y^2}dyP_{gg}(\frac{x}{y},\alpha_{s}(r))\sigma_{\mathrm{DGLAP}}(x,r).~~
\end{eqnarray}
Now, we rewrite $\sigma_{\mathrm{DGLAP}}$ in Eq.(13) in terms of
the variables $\upsilon=\ln(1/x)$ and $r$ instead of $x$ and $r$
by using the notation
$\widehat{\sigma}_{\mathrm{DGLAP}}(\upsilon,r){\equiv}{\sigma}_{\mathrm{DGLAP}}(e^{-\upsilon},r)$
as
\begin{eqnarray}
d\widehat{\sigma}_{\mathrm{DGLAP}}(\upsilon,r)=\bigg{[}\frac{2}{r}+\frac{d{\ln}\alpha_{s}(r)}{dr}\bigg{]}{dr}\widehat{\sigma}_{\mathrm{DGLAP}}(\upsilon,r)\nonumber\\
-\frac{2C}{r^3\bigg{(}\frac{C}{r^2}+\mu_{0}^{2}\bigg{)}}dr{\int_{0}^{\upsilon}}\bigg{[}e^{-(\upsilon-w)}\widehat{P}_{gg}(\upsilon-w,\alpha_{s}(r))\nonumber\\
{\times}\widehat{\sigma}_{\mathrm{DGLAP}}(w,r)\bigg{]}{dw}.~~~~~~
\end{eqnarray}
In the following, we use the Laplace transform method developed in
detail in \cite{Ref11, Ref12, Ref13, Ref14, Ref15} as
$\mathcal{L}[\widehat{\sigma}_{\mathrm{DGLAP}}(\upsilon,r);s]{\equiv}{\sigma}_{\mathrm{DGLAP}}(s,r)$
and using the fact that the Laplace transform of a convolution
factors is simply the ordinary product of the Laplace transform of
the factors, as
\begin{eqnarray}
\mathcal{L}\bigg{[}\int_{0}^{\upsilon}\widehat{\sigma}_{\mathrm{DGLAP}}(w,r)
\widehat{H}(\upsilon-w,\alpha_{s}(r))dw;s\bigg{]}\nonumber\\
={\sigma}_{\mathrm{sD}}(s,r){\times}h(s,\alpha_{s}(r)),~~
\end{eqnarray}
where
\begin{eqnarray}
h(s,\alpha_{s}(r)){\equiv}\mathcal{L}[e^{-\upsilon}\widehat{P}_{gg}(\upsilon,\alpha_{s}(r));s]=
\frac{\alpha_{s}(r)}{2\pi}h^{(0)}(s)\nonumber\\
+\bigg{(}\frac{\alpha_{s}(r)}{2\pi}\bigg{)}^2h^{(1)}(s),\hspace{3cm}
\end{eqnarray}
where
\begin{eqnarray}
h^{(0)}(s)=\frac{33-2n_{f}}{6}+6\bigg{(}\frac{1}{s}-\frac{2}{1+s}+\frac{1}{s+2}\nonumber\\
-\frac{1}{s+3}-\Psi(s+1)-\gamma_{E}\bigg{)},\hspace{1cm}
\end{eqnarray}
where $\Psi(x)$ is the digamma function and
$\gamma_{E}=0.5772156....$ is Euler$^{,}$s constant.\\
 Therefore, we find
\begin{eqnarray}
\sigma_{sD}(s,r)=\sigma_{sD}(s,r_{0})\bigg{(}\frac{r}{r_{0}}\bigg{)}^{2}\frac{\alpha_{s}(r)}{\alpha_{s}(r_{0})}{\times}\nonumber\\
\exp\bigg{[}-\frac{h^{(0)}(s)}{2\pi}\int_{r_{0}}^{r}\frac{2C\alpha_{s}(r)}{r^3\bigg{(}\frac{C}{r^2}+\mu_{0}^{2}\bigg{)}}dr\nonumber\\
-\frac{h^{(1)}(s)}{(2\pi)^{2}}\int_{r_{0}}^{r}\frac{2C\alpha^{2}_{s}(r)}{r^3\bigg{(}\frac{C}{r^2}+\mu_{0}^{2}\bigg{)}}dr\bigg{]}.
\end{eqnarray}
We keep the $1/s$ terms in the high-energy region of the
coefficients $h^{(0)}(s)$ and $h^{(1)}(s)$ and apply the inverse
Laplace transform as we find
\begin{eqnarray}
\widehat{\sigma}_{\mathrm{DGLAP}}(\upsilon,r){\simeq}\int_{0}^{\upsilon}
\widehat{\eta}(w,r,r_{0})\widehat{J}(\upsilon-w,\alpha_{s}(r))dw,
\end{eqnarray}
where
\begin{eqnarray}
\widehat{\eta}(\upsilon,r,r_{0})=\widehat{\sigma}_{\mathrm{DGLAP}}(\upsilon,r_{0})\bigg{(}\frac{r}{r_{0}}\bigg{)}^{2}\frac{\alpha_{s}(r)}{\alpha_{s}(r_{0})},
\end{eqnarray}
and
\begin{eqnarray}
\widehat{J}(\upsilon,\alpha_{s}(r))=\delta(\upsilon)+\frac{\sqrt{-\xi}}{\sqrt{\upsilon}}\mathrm{BesselI}(1,2\sqrt{-\xi}\sqrt{\upsilon}),
\end{eqnarray}
where
\begin{eqnarray}
\xi=\frac{2C_{A}}{2\pi}\int_{r_{0}}^{r}\frac{2C\alpha_{s}(r)}{r^3\bigg{(}\frac{C}{r^2}+\mu_{0}^{2}\bigg{)}}dr\hspace{2cm}\nonumber\\
+\frac{(12C_{F}T_{f}-46C_{A}T_{f})}{9(2\pi)^{2}}\int_{r_{0}}^{r}\frac{2C\alpha^{2}_{s}(r)}{r^3\bigg{(}\frac{C}{r^2}+\mu_{0}^{2}\bigg{)}}dr,
\end{eqnarray}
where $C_{F}=\frac{N_{c}^{2}-1}{2N_{c}}$ and $T_{f}=T_{R}n_{f}$.\\
 Transforming back in to $x$ space,
${\sigma}_{\mathrm{DGLAP}}(x,r)$ is given by
\begin{eqnarray}
{\sigma}_{\mathrm{DGLAP}}(x,r)=\bigg{(}\frac{r}{r_{0}}\bigg{)}^{2}\frac{\alpha_{s}(r)}{\alpha_{s}(r_{0})}\bigg{[}{\sigma}_{\mathrm{DGLAP}}(x,r_{0})\hspace{1cm}\nonumber\\
+\int_{x}^{1}{\sigma}_{\mathrm{DGLAP}}(z,r_{0})\frac{\sqrt{-\xi}}{\sqrt{{\ln}\frac{z}{x}}}\mathrm{BesselI}(1,2\sqrt{-\xi}\sqrt{{\ln}\frac{z}{x}})\frac{dz}{z}.~~
\end{eqnarray}
Therefore, the evolution of the color dipole cross section is
defined by the following form
\begin{eqnarray}
\sigma_{\mathrm{dip}}(x,r)=\sigma_{0}\{1-\exp(-\frac{1}{\sigma_{0}}\bigg{(}\frac{r}{r_{0}}\bigg{)}^{2}\frac{\alpha_{s}(r)}{\alpha_{s}(r_{0})}
\bigg{[}{\sigma}_{\mathrm{DGLAP}}({x},r_{0})\nonumber\\
+\int_{x}^{1}{\sigma}_{\mathrm{DGLAP}}(z,r_{0})\frac{\sqrt{-\xi}}{\sqrt{{\ln}\frac{z}{x}}}\mathrm{BesselI}(1,2\sqrt{-\xi}\sqrt{{\ln}\frac{z}{x}})\frac{dz}{z})\},~
\end{eqnarray}
where
\begin{eqnarray}
{\sigma}_{\mathrm{DGLAP}}(x,r_{0})=\frac{\pi^{2}}{N_{c}}r_{0}^{2}\alpha_{s}(r_{0})xg({x},r_{0}).
\end{eqnarray}
The initial gluon distribution at the scale $\mu_{0}^{2}$ is
defined in the form \cite{Ref2}
\begin{eqnarray}
xg(x,\mu_{0}^{2})=A_{g}x^{-\lambda_{g}}(1-x)^{C_{g}},
\end{eqnarray}
where parameters are motivated by global fits to DIS data with
respect to the results in Ref.\cite{Ref7}.\\
Evolution of the gluon distribution with respect to the
Laplace-transform method at low $x$ is defined by
\begin{eqnarray}
xg(x,\mu^2)&{\simeq}& xg(x,\mu^2_{0})+\int_{x}^{1}xg(z,\mu^2_{0})\frac{\sqrt{\eta}}{\sqrt{{\ln}\frac{z}{x}}}\nonumber\\
&&{\times}\mathrm{BesselI}(1,2\sqrt{\eta}\sqrt{{\ln}\frac{z}{x}})\frac{dz}{z})\},
\end{eqnarray}
where
\begin{eqnarray}
\eta=\int_{\mu^2_{0}}^{\mu^2}\bigg{[}\frac{3}{\pi}\alpha_{s}(\mu^2)
-\frac{61}{9\pi^{2}}\alpha^{2}_{s}(\mu^2)\bigg{]}d{\ln}{\mu^2}.
\end{eqnarray}
The dipole cross section via a Fourier transform is related to the
unintegrated gluon distribution (UGD) \cite{Ref1, Ref16}  as the
UGD\footnote{The different models \cite{Ref17, Ref18, Ref19,
Ref20, Ref21, Ref4} have been derived where relative the UGD to
the $k_{t}$
factorization and have been compared in Refs.\cite{Ref22, Ref23, Ref24, Ref25, Ref26, Ref27, Ref28} as:\\
$\bullet$ In the IN model \cite{Ref17} a UGD soft-hard model has
been
proposed in the large- and small- $k_{t}$ region.\\
$\bullet$ The ABIPSW model \cite{Ref18} is a $x$-independent UGD
model.\\
$\bullet$ The HSS model \cite{Ref19} is defined to take a
convolution form between the BFKL gluon Green$^{,}$s function and
the leading-order of the proton impact factor.\\
$\bullet$ The WMR model  \cite{Ref20} gives the probability of
evolving from the scale $k_{t}$ on an extra-scale $\mu$ at fixed
$Q$.\\
$\bullet$ The GBW model \cite{Ref21, Ref4} is derived from the
dipole cross section.\\
$\bullet$ The UGD model based on gluon TMDs is considered in
Refs.\cite{Ref27, Ref29}.}, $f(x,k_{t}^{2})$, is defined through
the differential integrated gluon distribution function by the
following form
\begin{eqnarray}
f(x,k_{t}^{2})=\frac{\partial{xg(x,\mu^2)}}{\partial{\ln}\mu^2}|_{\mu^2=k_{t}^{2}}
\end{eqnarray}
where the evolution of the UGD is find
\begin{eqnarray}
f(x,k_{t}^{2})&=&
\int_{x}^{1}xg(z,\mu^2_{0})\frac{\partial}{\partial{\ln}\mu^2}\bigg{[}\frac{\sqrt{\eta}}{\sqrt{{\ln}\frac{z}{x}}}\nonumber\\
&&{\times}\mathrm{BesselI}(1,2\sqrt{\eta}\sqrt{{\ln}\frac{z}{x}})\bigg{]}\frac{dz}{z})\}
\bigg{|}_{\mu^2=k_{t}^{2}}
\end{eqnarray}
Unintegrated distribution is modified by the Sudakov form factor
\cite{Ref30, Ref31} so that the UGD remains true as $x$ increases
or decreases \cite{Ref32, Ref33}. We find
\begin{eqnarray}
f^{S}(x,k_{t}^{2})&=&\bigg{[}xg(x,\mu_{0}^{2})\frac{\partial}{\partial{\ln}\mu^{2}}e^{-S(r,\mu^2)}\nonumber\\
&&+\int_{x}^{1}xg(z,\mu^2_{0})\frac{\partial}{\partial{\ln}\mu^2}\bigg{[}e^{-S(r,\mu^2)}\frac{\sqrt{\eta}}{\sqrt{{\ln}\frac{z}{x}}}\nonumber\\
&&{\times}\mathrm{BesselI}(1,2\sqrt{\eta}\sqrt{{\ln}\frac{z}{x}})\bigg{]}\frac{dz}{z})\}\bigg{]}
\bigg{|}_{\mu^2=k_{t}^{2}}
\end{eqnarray}
with
\begin{eqnarray}
S_{\mathrm{pert}}(r,\mu^2)&=&\frac{C_{A}}{2{\pi}b_{0}}\bigg{[}-\ln{\bigg{(}}\frac{\mu^{2}}{\mu_{b}^2}{\bigg{)}}
+{\bigg{(}}
\frac{1+\alpha_{s}(\mu^2_{b})b_{0}\ln{\bigg{(}}\frac{\mu^{2}}{\mu_{b}^2}{\bigg{)}}}{\alpha_{s}(\mu^2_{b})b_{0}}
{\bigg{)}}\nonumber\\
&&{\times}\ln{\bigg{(}}1+\alpha_{s}(\mu^2_{b})b_{0}\ln{\bigg{(}}\frac{\mu^{2}}{\mu_{b}^2}{\bigg{)}}{\bigg{)}}\bigg{]},
\end{eqnarray}
where $\mu_{b}=2e^{-\gamma_{E}}/r$ and $\gamma_{E}{\approx}0.577$
is the Euler-Mascheroni constant.\\

\subsection{3. Results}

In Fig.1, using the initial scale for the gluon distribution
$xg(x,\mu_{0}^{2})$ in Eq.(26) and Table I, we have plotted the
gluon distribution function, $xg(x,\mu^2)$, from Eq.(27) as a
function of $x$ (for $x<0.1$) for the values of $\mu^2=1.9, 10$
and $100~\mathrm{GeV}^2$ at the leading-order (LO) approximation
and compared with the CJ15 parametrization method \cite{Ref34} as
accompanied by the total errors. The gluon distribution extracted
at $\mu^2$ values are in good agreement with the CJ15 data. The
QCD parameter $\Lambda$ for four numbers of active flavor has been
extracted due to $\alpha_{s}(M_{z}^{2})=0.1166$ in comparison with
the CJ15 results with
$\alpha_{s}(M_{z}^{2})=0.136668$.\\
In Fig. 2 we show the logarithmic $k_{t}^{2}$-derivative of the
evolution of  gluon distribution as a function of $k_{t}^{2}$ for
three values of $x$. We observe that the function $f(x,k_{t}^{2})$
increases as $k_{t}^{2}$ increases and the variable $x$ decreases.
The  behavior  of UGD decreases as $k_{t}^{2}$ decreases to
$k_{t}^{2}<1~\mathrm{GeV}^2$. The resulting UGD with the
$k_{t}^{2}$ dependence from the evolution of the gluon
distribution and comparison with the HSS \cite{Ref19}, IN
\cite{Ref19} and WMR \cite{Ref20} models at $x=10^{-3}$ and
$x=10^{-4}$ are shown in Figs.3 and 4 respectively. We observe
that the results in Figs.3 and 4, due to the evolution of the
gluon distribution, are comparable to the models (especially
IN-CT14 LO ) at moderate $k_{t}^{2}$. The behavior of the UGD at
small and large $k_{t}^{2}$ in Figs.3 and 4 decreases and
increases in comparison with the models respectively. There is an
improved model (i.e., Eq.(31) ) due to the Sudakov form factor,
which increases and decreases the UGD behavior  at small and large
values of $k_{t}^{2}$ respectively. In Fig.5, we show the effects
of this correction in the ratio
$R_{f}=\frac{f^{S}(x,k_{t}^{2})}{f(x,k_{t}^{2})}$ for
$k_{t}^{2}{\geq}2~\mathrm{GeV}^{2}$.\\
\begin{figure}
\includegraphics[width=0.55\textwidth]{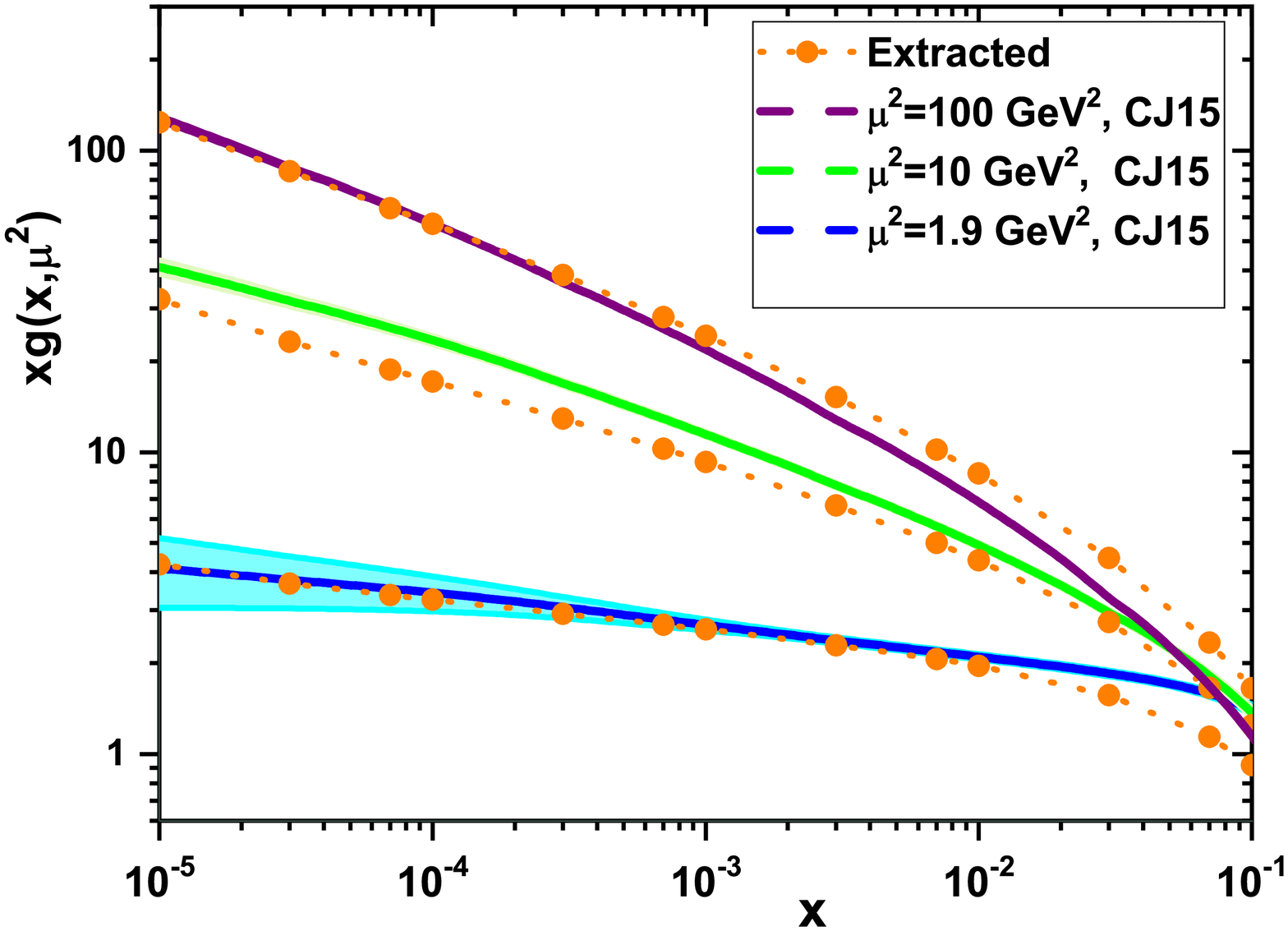}
\caption{The extracted $xg(x,\mu^2)$ as a function of $x$ compared
with the CJ15LO parametrization method \cite{Ref34} at
$\mu^2=1.9,~ 10~ \mathrm{and}~ 100~\mathrm{GeV}^2$.}\label{Fig1}
\end{figure}
\begin{table}
\centering \caption{The fixed parameters of the gluon density at
the initial scale  and dipole cross section
 according to Ref.\cite{Ref7} with $m_{c}=1.4~\mathrm{GeV}$.
  }\label{table:table1}
\begin{minipage}{\linewidth}
\renewcommand{\thefootnote}{\thempfootnote}
\centering
\begin{tabular}{|l|c|c|c|c|c|} \hline\noalign{\smallskip}
$\mu_{0}^{2}~[\mathrm{GeV}^2]$ & C & $\sigma_{0}~[\mathrm{mb}]$ & $A_{g}$  & $\lambda_{g}$ & $C_{g}$\\
\hline\noalign{\smallskip}
 1.73 & 0.38 & 22.40 & 1.35 & 0.079 & 5.6  \\
\hline\noalign{\smallskip}
\end{tabular}
\end{minipage}
\end{table}
\begin{figure}
\includegraphics[width=0.5\textwidth]{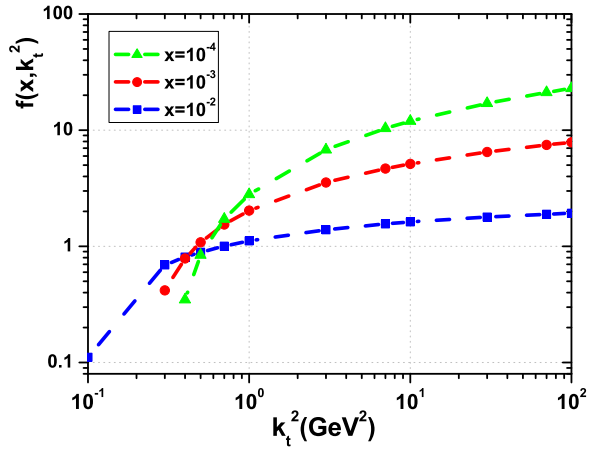}
\caption{The evolution of the unintegrated gluon distribution from
Eq.(30).}\label{Fig2}
\end{figure}
\begin{figure}
\includegraphics[width=0.5\textwidth]{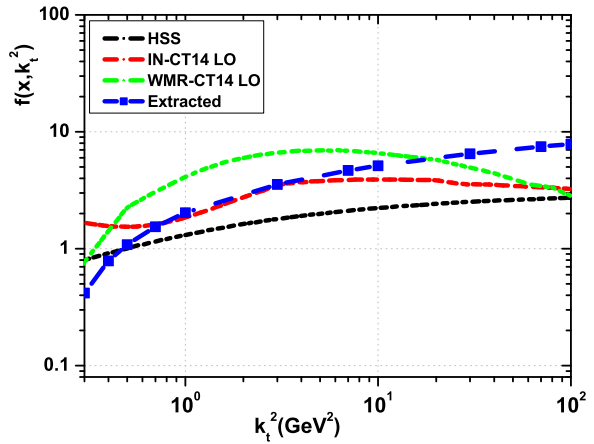}
\caption{UGD extracted (square-dash curve) from the evolution of
gluon distribution  compared with Hentschinski et al.
\cite{Ref19}, Ivanov and Nikolaev \cite{Ref17}, and Watt et al.
\cite{Ref20} models (dashed curves) as a function of $k_{t}^{2}$
at $x=10^{-3}$.}\label{Fig3}
\end{figure}
\begin{figure}
\includegraphics[width=0.5\textwidth]{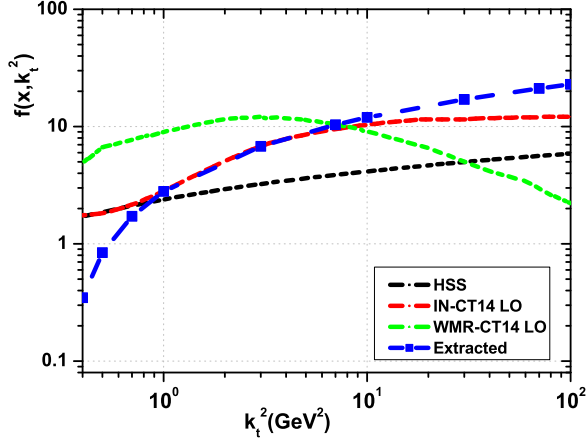}
\caption{The same as Fig.3 for $x=10^{-4}$.}\label{Fig4}
\end{figure}
\begin{figure}
\includegraphics[width=0.5\textwidth]{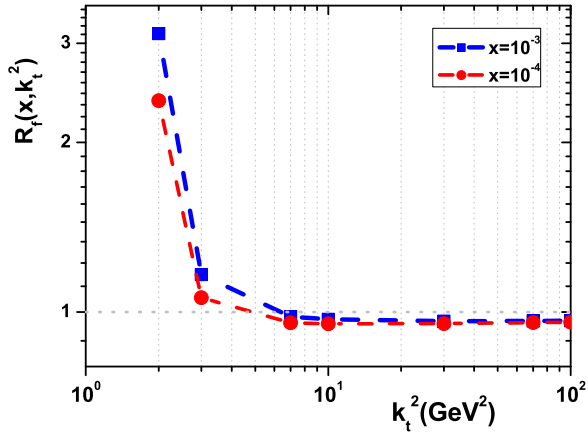}
\caption{The ratio of Sudakov form factor effect in the UGD for
$x=10^{-3}$ and $x=10^{-4}$ at
$k_{t}^{2}{\geq}2~\mathrm{GeV}^{2}$.}\label{Fig5}
\end{figure}
\begin{figure}
\includegraphics[width=0.5\textwidth]{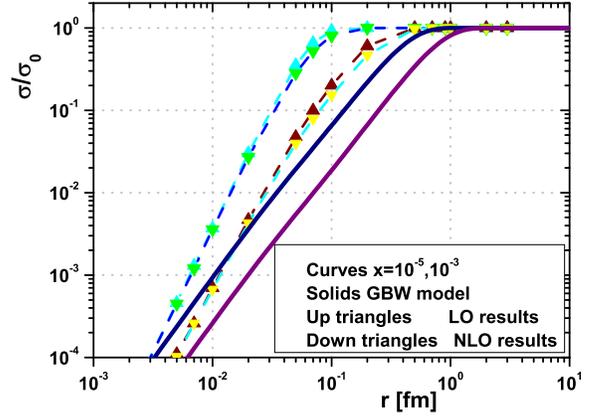}
\caption{The extracted ratio $\sigma_{\mathrm{dip}}/\sigma_{0}$ as
a function of $r$ for $x=10^{-5}$ and $10^{-3}$ (curves from left
to right, respectively) at the LO (up triangles) and NLO (down
triangles) approximations from the evolution of the dipole cross
section, Eq.(24), compared with the GBW model (solid curves),
Eq.(2).}\label{Fig1}
\end{figure}
\begin{figure}
\includegraphics[width=0.5\textwidth]{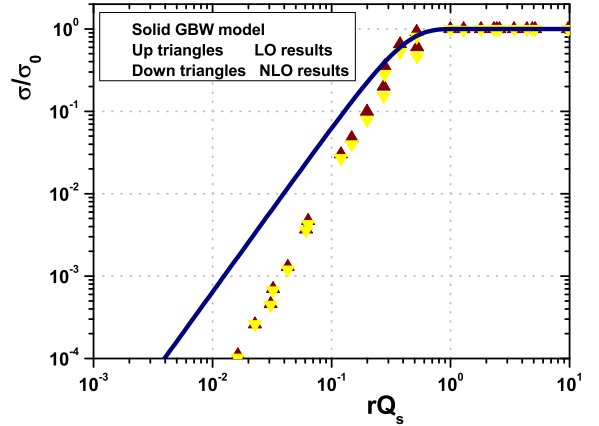}
\caption{ The extracted ratio
$\sigma_{\mathrm{dip}}(rQ_{s}(x))/\sigma_{0}$ as a function of
$rQ_{s}(x)$ at the LO (up triangles) and NLO (down triangles)
approximations from the evolution of the dipole cross section,
merges into one line due to the geometric scaling compared with
the GBW model (solid curve).}\label{Fig2}
\end{figure}
We have calculated the $r$ dependence of the ratio
$\sigma_{\mathrm{dip}}/\sigma_{0}$ for $x=10^{-5}$ and $10^{-3}$
at the LO  and NLO  approximations\footnote{The charm quark mass
is taken to account in this analysis.} from the evolution of the
dipole cross section, Eq.(24), and compared with the GBW model,
Eq.(2), in Fig.6. The evolution of the dipole cross section gives
a similar description of the ratio
$\sigma_{\mathrm{dip}}/\sigma_{0}$ in comparison with the GBW
model at low $x$. The transition point between the large and small
$r$ regimes\footnote{At small $r$, $\sigma_{\mathrm{dip}}$
features colour transparency which is pQCD phenomenon and  for
large $r$ it saturates \cite{Ref4}.} in the evolution model (i.e.,
Eq.(24)) increases to the lower $r$ values in comparison with the
GBW model, although is governed by the increasing with $x$
saturation radius $R_{0}(x){\equiv}\frac{1}{Q_{s}(x)}$. The
evolution model is guaranteed that the dipole cross section
saturates for smaller dipole size in comparison with the GBW model
and decreasing $x$.\\
The evolution of the dipole cross section (i.e., Eq.(24)) has a
property of geometric scaling as it becomes a function of a single
variable, $rQ_{s}$, for all values of $r$ and $x$. In Fig.7, we
show the ratio $\sigma_{\mathrm{dip}}(rQ_{s}(x))/\sigma_{0}$ in
the evolution method in the single variable $rQ_{s}(x)$ for all
values of $x$ and $r$ at the LO and NLO approximations. The
results at $x=10^{-5}$ and $10^{-3}$ merge into one solid line
where this is a reflecting of geometrical scaling. This
geometrical scaling is visible in the region $rQ_{s}(x){\geq}0.2$
together with its violation for $rQ_{s}(x)<0.2$ in comparison with
the GBW model.\\

In conclusion, we have presented a method based on the Laplace
transform method to determine the evolution of the dipole cross
section at the LO and NLO approximations. This method relies on
the BGK method within a kinematical region characterized by low
values of the Bjorken variable $x$ in a wide range of transverse
size $r$. In the first part of this analysis, we obtained a new
evolution for the integrated and unintegrated gluon distributions
in a wide range of $\mu^2$ and $k_{t}^{2}$ respectively. We have
shown that the evolution of integrated gluon distribution is
comparable with the CJ15 parametrization group in a wide range of
$\mu^2$ and $x$ and also the evolution of unintegrated gluon
distribution  is comparable with the UGD models in a wide range of
$k_{t}^{2}$. The effects of the Sudakov form factor were
investigated for the UGD in a wide range of $k_{t}^{2}$. The
effect is visible with a small and large value of $k_{t}^{2}$. In
the second part of our study, we have compared the evolution of
the color dipole cross section with the GBW model in a wide range
of $r$ at low values of $x$. This evolution method allows to
extend the GBW saturation model to low values of dipole size $r$
(or large values of $Q^2$) and shows that the saturation scale and
geometric scaling are retained in the region
$rQ_{s}(x){\geq}0.2$.\\

\subsection{ACKNOWLEDGMENTS}

The author is thankful to Razi university for financial support of
this project and  would like to thank Professor Jen-Chieh Peng and
B. Z. Kopeliovich for useful comments.


\end{document}